\documentclass[10pt,aps,showpacs,eqsecnum]{revtex4}

\usepackage{graphicx,color}
\usepackage{amssymb}

\mathchardef\bigtilde="0365

\begin{document}

\title{Critical behaviors as functions of the bare-mass}
\author{Hirofumi Yamada}\email{yamada.hirofumi@it-chiba.ac.jp}
\affiliation{%
Division of Mathematics and Science, Chiba Institute of Technology, 
\\Shibazono 2-1-1, Narashino, Chiba 275-0023, Japan}



\date{\today}

\begin{abstract}
{In Ising model on the simple cubic lattice, we describe the inverse temperature $\beta$ in terms of the bare-mass $M$ and study its critical behavior by the use of delta expansion from high temperature or large $M$ side.   In the vicinity of critical temperature $\beta_{c}$, the expansion of $\beta$ in $M$ has $\beta_{c}$ as the first term and $M^{-1/2\nu}$ as the leading correction.   The estimation of $\beta_{c}$ in $1/M$ expansion is confronted with the leading and higher order corrections, even delta expansion is applied and the critical region emerges.    To improve the estimation status of $\beta_{c}$,  we try to suppress the corrections by adding derivatives of $\beta(M)$ with free adjustable parameters.  By optimizing the parameters with the help of the principle of minimum sensitivity which are maximally imposed in accord with the number of parameters, estimation of $\beta_{c}$ is carried out and the result is found to be in good agreement with the present world average.  In the same time, the critical exponent $\nu$ is also estimated.
}
\end{abstract}


\maketitle

\section{Introduction}
Since the invention of lattice field theories, the border between condensed matter models and field theoretic models is lost and the techniques in the statistical physics have been frequently used in the field theory analysis~\cite{wil,kog}.  Traditionary, the both systems are described in terms of $\beta$ which indicates inverse temperature or inverse bare coupling constant.   In the field theory side, we however notice that the lattice spacing $a$ or the equivalent bare-mass can play the role of the basic parameter describing the models.  This feature naturally appears in the large $N$ limit of field theoretic models.    Also for finite-$N$ case of non-linear sigma models, it was shown that the scaling behavior of the inverse bare coupling has been captured from large bare-mass expansion~\cite{yam}.   Then, it is a small and natural step to take the reverse point of view into the condensed matter models.   The motivation of this work is to investigate whether such reverse approach to the critical phenomena is effective or not.

In the present paper, we concern with the 3-dimensional Ising model on the simple cubic lattice as a theoretical laboratory.  
Let us sketch our strategy below:  As the temperature approaches to the critical one from the high temperature side, the correlation length $\xi$ diverges as
\begin{equation}
\xi\to f_{+}(1-\beta/\beta_{c})^{-\nu}\{1+const\times(1-\beta/\beta_{c})^{\theta}+\cdots\} ,\quad \beta<\beta_{c},
\label{xi}
\end{equation}
where $\beta_{c}$, $\nu$, $f_{+}$ and $\theta$ stand for the critical temperature, the exponent associated with $\xi$,  the amplitude in the high temperature phase and the exponent of confluent singularity~\cite{weg}, respectively.  
The scaling law (\ref{xi}) can be rewritten in terms of the bare-mass $M$, where $M$ is defined by the magnetic susceptibility $\chi$ and the second moment $\mu$ as
\begin{equation}
M=\frac{2D\chi}{\mu},\quad D: {\rm space\,\, dimension}.
\label{mass}
\end{equation}
First we invert (\ref{xi}) and obtain $\beta=\beta_{c}(1-f_{+}^{1/\nu} \xi^{-\frac{1}{\nu}}+\cdots)$.  Then, since $\xi\sim M^{-1/2}$ in the critical region, we have
\begin{equation}
\beta=\beta_{c}-\beta_{c}f_{+}^{1/\nu} M^{\frac{1}{2\nu}}+\cdots,
\label{betamass}
\end{equation}
where the "$\cdots$" represents the higher order corrections.  Thus the critical temperature is given by the limit 
\begin{equation}
\beta_{c}=\lim_{M\to 0}\beta(M),
\label{limit}
\end{equation}
and $(2\nu)^{-1}$ in (\ref{betamass}) is interpreted as the exponent of the leading correction.  The description by $M$ is not restricted to $\beta$ but is applicable to $\chi$~\cite{yam1}, specific heat and maybe others.  For instance, we can express $\chi$ by $M$ via the relation with $\beta$ such that 
$\chi\to const\times M^{-\gamma/(2\nu)}$ as $M\to 0$.  
Note that we need no information of $\beta_{c}$ to study the small $M$ behavior of $\chi(M)$ and the estimation of $\gamma$ is unbiased.   

In the study of the scaling behavior of $\beta$, we confine with the high temperature phase and use the $1/M$ expansion.  Apparently, we need some improvement to allow the use of the series near the critical point.  As a key technique, we apply the so-called delta expansion method on the lattice~\cite{yam2}.  Actually in~\cite{yam2}, the Ising model on the square lattice was revisited to examine the power of the method.  Though a remarkable improvement on the behavior of high temperature expansion was shown, the discussion there ended at the semi-quantitative level and the estimation task of $\beta_{c}$ and $\nu$ was not attempted.  The difficulty for the accurate quantitative study of critical quantities comes from the corrections to the asymptotic scaling, mainly from the second term in (\ref{betamass}) and even from higher orders.  In the present paper, we tackle the problem by introducing freely adjustable parameters into the naive thermodynamic quantities.  For $\beta$, we instead consider
\begin{equation}
\psi(\{\rho_{k}\},x)=\Big[\sum_{k=0}\rho_{k}\Big(x\frac{d}{dx}\Big)^{k}\Big]\,\beta(x),\quad x=M^{-1}.
\end{equation} 
Here, $\rho_{0}=1$ and $ \rho_{1}=\rho, \,\rho_{2}=\sigma, \,\cdots$ are adjustable free parameters.   By exploiting a special property of the delta expanded $\psi$, we make approximate cancellation of corrections to $\beta_{c}$ by seeking optimal values of parameters order by order.  
   On the basis of the high temperature or large bare-mass expansion, we then try to compute $\beta_{c}$ and $\nu$.  The values of $\beta_{c}$ and $\nu$ have been computed by various methods including Monte Carlo simulation, field theoretic methods centered on the renormalization group and series expansions (See, for a review~\cite{peli} and recent researches~\cite{ari,den,ber,pog,has,lit,gor}).  The best estimation of $\beta_{c}$ known to our knowledge is $\beta_{c}= 0.2216546(10)$~\cite{blo} and $\beta_{c}= 0.2216595(15)$~\cite{ito}.  In this work, we however quote modest one
\begin{equation}
\beta_{c}=0.22165,
\label{betac}
\end{equation}
to which all recent literatures agree up to the last digit.  For $\nu$ we refer
\begin{equation}
\nu=0.630,
\label{nu}
\end{equation}
which is also agreed by all recent works within $10^{-3}$ order.
   
This paper is organized as follows:  In the next section, we briefly review the method of delta expansion on the lattice.   In the third section, we attempt to estimate critical quantities, $\beta_{c}$ and $\nu$, of the Ising model on the simple cubic lattice by the use of the delta expansion on $\psi$.    Studies on the magnetic susceptibility and specific heat, including also for the square lattice model as a bench mark of the method, are now under the progress.  To avoid a report too long, we hope to carry those studies forward to another publication.  In the present paper, we like to present the essence of the idea and technical details of our approach by focusing on the relation of $\beta$ and $M$  on the cubic lattice.   The conclusion is stated in the last section.

\section{Delta expansion}
To access the critical region, we make an attempt to dilate the region itself around $M=0$.  Given a thermodynamic quantity $f(M)$, we consider the dilated function $\bar f$, $\bar f(M,\delta)=f(M(1-\delta))$ with $0\le \delta\le 1$.   Setting the value of $\delta$ close to $1$, the critical region, the neighborhood of $M=0$, is enlarged to the region far from the origin.  Then, if the large $M$ series of $f(M(1-\delta))$ still effective in some region of $M$ can be available in the $\delta\to 1$ limit, the series may recover the original critical behavior within there.   To obtain such an effective large $M$ series of $\bar f(M,\delta)$, we have found that a protocol of obtaining best large $M$ series for $\bar f(M,\delta)$ is to treat $M^{-1}$ and $\delta$ on equal footing:  Let $M$ in $\bar f(M,\delta)$ denote $t^{-1}$ to avoid notational confusion.   Then consider the truncated series $f_{N}(M)=\sum_{n=0}^{N} a_{n}M^{-n}$ and its dilation, $f_{N}(t^{-1}(1-\delta))=\sum_{n=0}^{N} a_{n}\{t/(1-\delta)\}^{n}$.   If the full order of the series is $N$, the term $\{t/(1-\delta)\}^n$ should be expanded in $\delta$ and truncated such that the sum of orders $i$ and $j$ is equal to or less than $N$ where $i$ and $j$ denote respectively the orders of $t$ and $\delta$~\cite{yam2,yam3}.  In this rule, $\{t/(1-\delta)\}^n$ should be expanded in $\delta$ up to the order $\delta^{N-n}$, and we find
\begin{equation}
\{t/(1-\delta)\}^n\sim t^{n}\Big(1+n\delta+\frac{n(n+1)}{2!}\delta^2+\cdots+\frac{n(n+1)\cdots(n+N-n-1)}{(N-n)!}\delta^{N-n}\Big).
\end{equation}
After the expansion in $\delta$, we can take the limit $\delta\to1$.  The result gives the transform,
\begin{equation}
\{t/(1-\delta)\}^n\to C(N,n)t^{n},
\label{prescription}
\end{equation}
where
\begin{equation}
C(N, n)=\frac{N!}{n!(N-n)!}=\frac{\Gamma(N+1)}{\Gamma(n+1)\Gamma(N-n+1)}.
\end{equation}
The above transformation rule is very simple.  If one has the truncated $1/M$ series of $f(M)$ to order $N$, one obtains readily the corresponding delta-expanded series.  That is, given a truncated series,
\begin{equation}
f_{N}=\sum_{n=0}^{N}a_{n}M^{-n},
\end{equation}
the corresponding delta-expanded series $D[f_{N}]$ reads
\begin{equation}
D[f_{N}]=\sum_{n=0}^{N}a_{n}C(N, n)t^n=\bar f_{N}(t).
\label{deltatr}
\end{equation}
We notice that $C(N, 0)=1$ and the constant term is left invariant.  

To summarize, the delta expansion creates a new function $\bar f_{N}(t)$ associated with $f_{N}(M)$ by the transform of the coefficient from $a_{n}$ to $a_{n}C(N, n)$ which depends on the order $N$.  The symbol $D$ denotes the transformation of $f_{N}(M)$ to $\bar f_{N}(t)$.   Since $C(N, n)\to \frac{N^n}{n!}$ ($n$:fixed, $N\to \infty$), one might think that the series becomes ill-defined in the $N\to \infty$ limit.  However, in some physical models and mathematical examples, we found the evidence that within some region, the limit indeed exists.    Consider, for example, the function $f(M)=(1+M)^{-1}=\sum_{n=1}^{\infty}(-1)^{n-1}M^{-n}$.  We find $\bar f_{N}(t)=\sum_{n=1}^{N}(-1)^{n-1}\frac{N!}{n!(N-n)!}t^n=1-(1-t)^N$ and the resulting function converges to $1$ for $|1-t|<1$ and diverges for $|1-t|>1$.   The point is, within the region $|1-t|<1$, $\lim_{N\to \infty}\bar f_{N}(t)\to 1=f(M=0)$.  Also from general point of view, the result is reasonable, since formally $\bar f(M, 1)=f(M=0)$.  In any way, in the present work, we assume that, over some region of $t$, the limit of the function sequence $\{f_{N}(t)\}$ tends to a constant one and
\begin{equation}
\lim_{N\to \infty}\bar f_{N}(t)=f(M=0).
\label{clue}
\end{equation}

In the presence of phase transition, we must deal with the case where the expansion of $f$ around $M=0$ is not regular.  Then we consider how in such a case the delta expansion affects the small $M$ behavior of $f(M)$,  supposing that $f$ behaves at small enough $M$ as $f\sim f(0)+f_{1}M^{\alpha_{1}}+f_{2}M^{\alpha_{2}}+\cdots$ where $0<\alpha_{1}<\alpha_{2}<\cdots$.  When $t^{-1}(1-\delta)$ is substituted into $M^{\alpha}$ and $((1-\delta)/t)^{\alpha}$ is expanded in $\delta$, giving
$M^{\alpha}=t^{-\alpha}(1-\alpha \delta+\cdots)$, 
a reasonable truncation protocol for the best matching with $\bar f_{N}(t)$ is not found on logical grounds.   Here also, we proceed along with experiences.  In some physical models, we found that the formal extension of  (\ref{prescription}), 
\begin{equation}
M^{\alpha}\to C(N, -\alpha)t^{-\alpha},\quad C(N, -\alpha)=\frac{\Gamma(N+1)}{\Gamma(-\alpha+1)\Gamma(N+\alpha+1)},
\label{ansatz}
\end{equation}
provides us the best matching.   The factor $C(N, -\alpha)$ vanishes when $\alpha=1,2,3,\cdots$ and for positive non-integer case goes to zero as
\begin{equation}
C(N, -\alpha)\to N^{-\alpha}/\Gamma(-\alpha+1)\to 0,\quad  (N\to \infty).
\label{prop1}
\end{equation}
Hence, the term of positive integer power of $M$ vanishes,
\begin{equation}
D[M^{\alpha}]=0,\quad (\alpha=1,2,3,\cdots),
\label{prop2}
\end{equation}
and the terms of fractional positive power of $M$ decreases with the order and disappears in the $N\to \infty$ limit.    Thus, we define the result of delta expansion for $M^{\alpha}$ by (\ref{ansatz}) and
\begin{equation}
D[f]=f(0)+f_{1}C(N, -\alpha_{1}) t^{-\alpha_{1}}+\cdots,\quad (t\gg 1).
\end{equation}
The above two results (\ref{prop1}) and (\ref{prop2}) show the main advantages of the delta expansion.  From these we understand that the approach to the critical behavior is quicker in the $\delta$-expanded function than the original function. 

In the present work, our task is to estimate $f(0)$ and $\alpha_{1}$ from the known series (\ref{deltatr}).  In the process, we use derivatives of $f_{N}$.  Then we remark that
\begin{equation}
t\frac{d}{dt}D[f_{N}]=D\Big[x\frac{d}{dx}f_{N}\Big],
\end{equation}
which states that $D$-operation and differentiation is commutable.  It is convenient to use the following abbreviate notation,
\begin{equation}
\Big(t\frac{d}{dt}\Big)^{k}\bar f_{N}=\bar f_{N}^{(k)},\quad (k=1,2,3,\cdots).
\end{equation}

\section{Estimation of $\beta_{c}$ and $\nu$}

\subsection{Preliminary study}
The Ising model on the simple cubic lattice is defined by the action
\begin{equation}
S=-\beta \sum_{< i,j>} s_{i}s_{j},\quad s_{i}^2=1,
\end{equation}
where the spin sum is over all nearest neighbour pair on the periodic lattice.  
Our approach is based upon the high temperature expansion.  The magnetic susceptibility and the second moment have been computed up to $\beta^{25}$ by Butera and Comi~\cite{butera}.  From (\ref{mass}) and the result reported in~\cite{butera}, we have
\begin{eqnarray}
\beta&=&\frac{1}{M}-\frac{6}{M^2}+\frac{124}{3M^3}-\frac{312}{M^4}+\frac{12596}{5M^5}-\frac{21432}{M^6}+\frac{1330848}{7M^7}-\frac{1745344}{M^8}\nonumber\\
& &+\frac{148384348}{9M^9}-\frac{797787336}{5M^{10}}+\frac{17341288504}{11M^{11}}-\frac{15857888272}{M^{12}}\nonumber\\
& &+\frac{2106367479672}{13M^{13}}-\frac{11748802870160}{7M^{14}}+\frac{263968267347944}{15M^{15}}\nonumber\\
& &-\frac{186504592354608}{M^{16}}+\frac{33924951987330804}{17M^{17}}-\frac{21535692193295224}{M^{18}}\nonumber\\
& &+\frac{4449606807205690200}{19M^{19}}-\frac{12821205881021198992}{5M^{20}}\nonumber\\
& &+\frac{197756701920466780928}{7M^{21}}-\frac{3442869826889278353376}{11M^{22}}\nonumber\\
& &+\frac{80156432259652309452520}{23M^{23}}-\frac{116948936021276297965072}{3M^{24}}\nonumber\\
& &+\frac{10946582972904015563857296}{25M^{25}}+O(M^{-26})\nonumber\\
&=&\sum_{n=1}^{\infty}\frac{b_{n}}{M^n}.
\label{beta_M}
\end{eqnarray}
The result of the delta expansion to the order $N$ is readily obtained by multiplying $n$th order coefficient by $C(N,n)$, giving
\begin{equation}
\bar\beta_{N}:=D[\beta_{N}]=\sum_{n=1}^{N}b_{n}C(N,n)t^n.
\label{delbeta}
\end{equation}

The effects of delta expansion are clearly shown in the plots of relevant functions.  
We show in Figure 1 the plots of $\beta_{25}$, $\bar\beta_{25}$, $\beta_{25}^{(1)}$, $\bar\beta_{25}^{(1)}$, $\beta_{25}^{(2)}$ and $\bar\beta_{25}^{(2)}$.    In the first graph, it is implied that $\bar\beta_{25}$ approaches to the correct $\beta_{c}$ (This is not the case for square Ising model).   Also in the derivatives, impressive point is demonstrated:   Beyond the peak about $t\sim 0.01$, $\bar\beta_{25}^{(1)}$ shows the monotonic decreasing trend to $t\sim 0.15$.  We have numerically checked by using the values, $f_{+}\sim 0.5$ for the amplitude~\cite{butera2} in (\ref{betamass}) and $\nu=0.63$ (see(\ref{nu})), the rough agreement of the  behaviors between $\bar\beta_{25}^{(1)}$ in $t$-series and its critical behavior $\bar\beta^{(1)}\sim \frac{f_{+}^{1/\nu}}{2\nu}C(25, -1/2\nu) t^{-1/2\nu}$ $(C(25, -1/2\nu)=0.0170152)$.  Same thing  applies to the second derivatives.   In addition, in the case of square Ising model, we ascertained by using known results of $\nu$ and the amplitude that the transformed derivatives exhibit critical behaviors.  Thus, we come to conclude that three plots in Figure 1 afford evidences emphasized in~\cite{yam2} that the delta expansion dilates the scaling region to the region of $t$ far from the infinity.
\begin{figure}[h]
\centering
\includegraphics[scale=0.5]{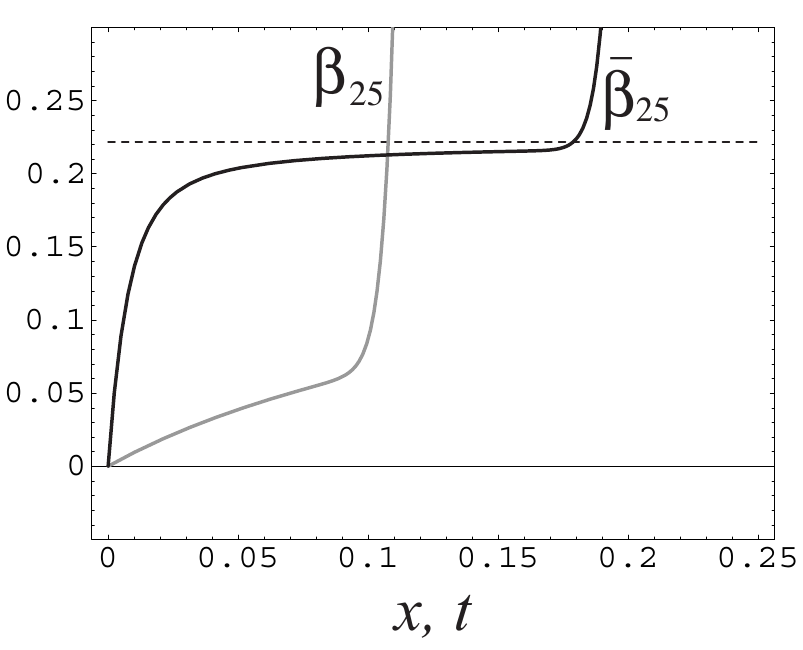}
\includegraphics[scale=0.5]{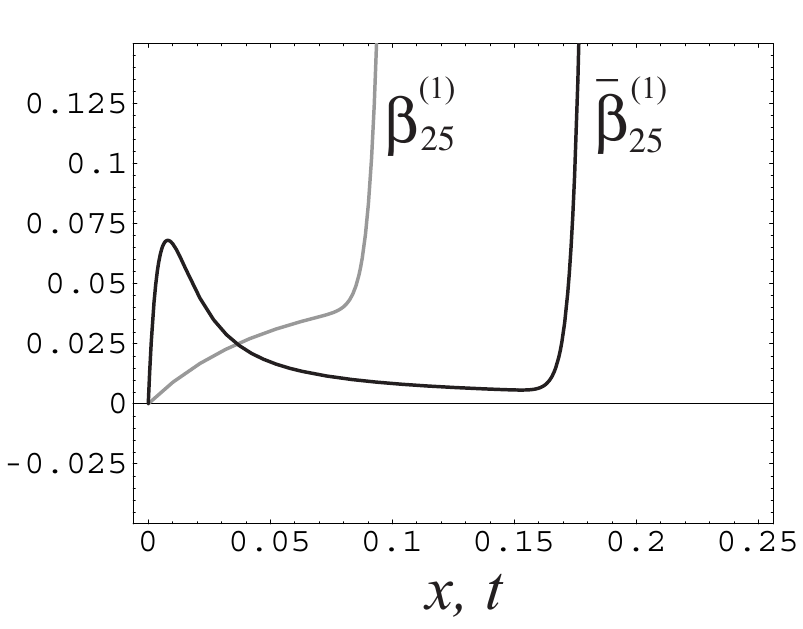}
\includegraphics[scale=0.5]{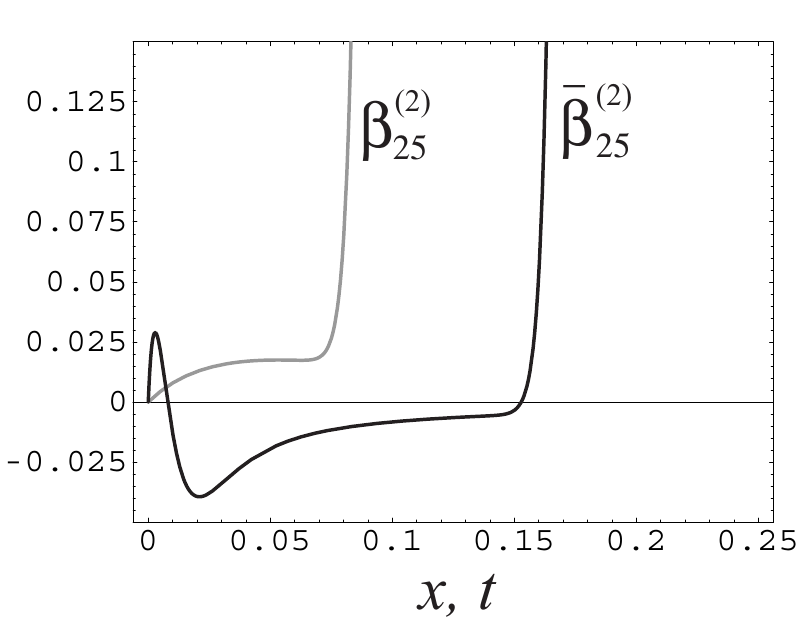}
\caption{Plots of $\beta_{25}$, $\bar\beta_{25}$, $\beta_{25}^{(1)}$, $\bar\beta_{25}^{(1)}$, $\beta_{25}^{(2)}$ and $\bar\beta_{25}^{(2)}$.  $\beta_{25}^{(k)}$ is shown in the gray plot and its delta expanded version $\bar\beta_{25}^{(k)}$ in the black plot.   In the first plot, the dotted line indicates the critical temperature $\beta_{c}=0.22165\cdots$ which value is widely confirmed.   In $\bar\beta_{25}$, the high temperature region is pushed away to the neighbourhood of the origin $t=0$ and the scaling region appears to cover up to point $t\sim 0.15$.  In the second plot, we see, though $\beta_{25}^{(1)}$ is monotonic, $\bar\beta_{25}^{(1)}$ has a peak around $t\sim 0.01$.  This peak indicates the turning point from high temperature region to the scaling region.  The same reasoning applies to $\bar\beta_{25}^{(2)}$ in the third plot.}
\end{figure}

In the study of phase transition via series expansion technique, we rely upon heuristic assumption of the power law or logarithmic behaviors near transition point.  
Let us start the argument by supposing the scaling behavior of inverse temperature in general form,
\begin{equation}
\beta=\beta_{c}-A_{1} x^{-p_{1}}-A_{2} x^{-p_{2}}-A_{3} x^{-p_{3}}-O(x^{-p_{4}}), \quad x=M^{-1},
\label{betascaling}
\end{equation}
where $0<p_{1}<p_{2}<p_{3}<\cdots$ and
\begin{equation}
\bar\beta=\beta_{c} -A_{1}C(N,-p_{1}) t^{-p_{1}}-A_{2}C(N,-p_{2}) t^{-p_{2}}-O(t^{-p_{3}}).
\label{delbetascaling}
\end{equation}
The comparison of (\ref{betascaling}) with (\ref{betamass}) gives
\begin{equation}
p_{1}=\frac{1}{2\nu}
\end{equation}
and the exponents of confluent singular terms ~\cite{weg} enter into $p_{i}$ where $i\ge 2$.   For instance, $p_{2}=(1+\theta)/2\nu$.    
Our assumption is that over some region of $t$, 
\begin{equation}
\lim_{N\to \infty}\bar\beta_{N}=\beta_{c}.
\label{clue2}
\end{equation}

Though the definition of $\beta_{c}$ is given by (\ref{limit}), it cannot be naively used for its estimation, because our information is limited to the truncated version (\ref{beta_M}) (One might think that Pad\`{e} approximants of (\ref{beta_M}) may be useful for the purpose.  Investigation on this direction is out of this work).  Moreover, though the behavior of $\bar\beta_{N}$ is remarkably improved as shown in the first plot in Figure 1, it is not enough yet to yield accurate estimation of $\beta_{c}$.  The reason is that the leading correction to $\beta_{c}$ given as $A_{1}C(N,-p_{1})t^{-p_{1}}$ is still active in the effective region of $\bar \beta_{N}$ for $N\le 25$.  Now then, our idea upon the estimation of $\beta_{c}$ is to cancel dominant  corrections to $\beta_{c}$ in (\ref{delbetascaling}) by introducing following $\psi$ with a free parameter $\rho$,
\begin{equation}
\psi(\rho, x)=\beta+\rho x\frac{d \beta}{dx}.
\end{equation}
At large $x$ (small $M$), $\psi(\rho, x)$ reads
\begin{equation}
\psi(\rho, x)=\beta_{c}-\sum_{n=1}^{\infty}(1-\rho p_{n})A_{n}x^{-p_{n}}.
\label{dif1}
\end{equation}
Note that for any value of $\rho$,
\begin{equation}
\lim_{x\to \infty}\psi(\rho, x)=\beta_{c}.
\label{psilimit}
\end{equation}
Then, we deal with $\delta$-expanded function 
\begin{equation}
\bar\psi_{N}(\rho, t)=D[\psi_{N}(\rho, x)]=\bar\beta_{N}+\rho\bar\beta_{N}^{(1)},
\end{equation} 
where $\psi_{N}(\rho, x)$ denotes the truncated series to the order $x^N$.  
It follows from (\ref{clue2}) that
\begin{equation}
\lim_{N\to \infty}\bar\psi_{N}(\rho, t)=\beta_{c}.
\label{delpsilimit}
\end{equation}
This should hold at any $t$ in the supposed convergent region.  We notice that, as the extension,  one can build $\psi$ with multi-parameters.   Making use of the property (\ref{delpsilimit}), we carry out estimation of $\beta_{c}$ and $\nu$ in the following subsections.

\subsection{$\beta_{c}$ and $\nu$ in one-parameter case}
\label{sec:one-para}
 When dealing with $\psi$, the critical issue is how we should choose $\rho$ value to cancel the correction in (\ref{delbetascaling}).  The starting point is (\ref{delpsilimit}) which states that $\bar\psi_{N}(\rho, t)$ approaches as $N\to \infty$ to a constant function over the convergent region.  First, at finite $N$, we employ the principle of minimum sensitivity due to Stevenson~\cite{steve}.    From (\ref{delpsilimit}),  the use of the principle is quite natural.  Thus, we postulate that, at a given $\rho$, the value of $\bar\psi_{N}(\rho, t)$ at the possible stationary point gives an estimation of $\beta_{c}$.  The resulting $\beta_{c}$ depends on $\rho$.  To extract best estimation, we notice that if the suppression of the correction is achieved, the coefficient of the first order correction in (\ref{dif1}) may almost vanish, leading $\rho\sim p_{1}^{-1}=2\nu$.  And then the stationary property of $\bar\psi_{N}$ becomes maximal.  Thus, taking reverse point of view, we search $\rho$ which realizes the maximal stationarity and minimization of the second derivative.  

The explicit procedure goes as follows: At a given $\rho$, the estimation of $\beta_{c}$ is achieved at $t$ satisfying
\begin{equation}
\bar\psi_{N}^{(1)}(\rho, t)=\bar\beta_{N}^{(1)}+\rho \bar\beta_{N}^{(2)}=0.
\label{rho}
\end{equation}
Then, to realize maximal stationarity, 
we further impose that optimal $\rho$ should locally minimize the absolute value of the second derivative $|\bar\psi_{N}^{(2)}|$ at $t$ which satisfies (\ref{rho}).  Then, our task is to try to adjust $\rho$ for the solution $t$ of (\ref{rho}) to become also the solution of the equation,
\begin{equation}
\bar\psi_{N}^{(2)}(\rho, t)=\bar\beta_{N}^{(2)}+\rho \bar\beta_{N}^{(3)}=0,
\label{rho2}
\end{equation}
or to make $|\bar\psi_{N}^{(2)}|$ locally minimum.  
The two conditions, (\ref{rho}) and being local minimum of $|\bar\psi_{N}^{(2)}|$, determine optimal $\rho=\rho^{*}$ and $t=t^{*}$ at which the stationarity is maximally realized.  In general, for a given $\rho$, the first condition (\ref{rho}) gives non-unique solutions of $t$ and it is convenient to express $\rho$ as a function of $t$, $\rho=\rho(t)$.   From (\ref{rho}), it follows that
\begin{equation}
\rho=-\frac{\bar\beta_{N}^{(1)}}{ \bar\beta_{N}^{(2)}},
\label{rho3}
\end{equation}
and then
\begin{equation}
\bar\psi_{N}^{(2)}(\rho, t)=((\bar\beta_{N}^{(2)})^2-\bar\beta_{N}^{(1)}\bar\beta_{N}^{(3)})/\bar\beta_{N}^{(2)}.
\label{cond1}
\end{equation}
With obtained $t^{*}$ and  $\rho^{*}=\rho(t^{*})$, we estimate $\beta_{c}$ by 
\begin{equation}
\beta_{c}(N)= \bar\psi_{N}(\rho^{*}, t^{*}).
\end{equation}

We carry out this procedure from $4$th order to $25$th order.  The order from which the real feature of our approach begins to show is $13$th order.   Here we mean by "real feature" the feature that the characteristic scaling behavior that the second derivative $\bar\psi_{N}^{(2)}$ should exhibit, which we describe below:
\begin{figure}[ht]
\centering
\includegraphics[scale=0.65]{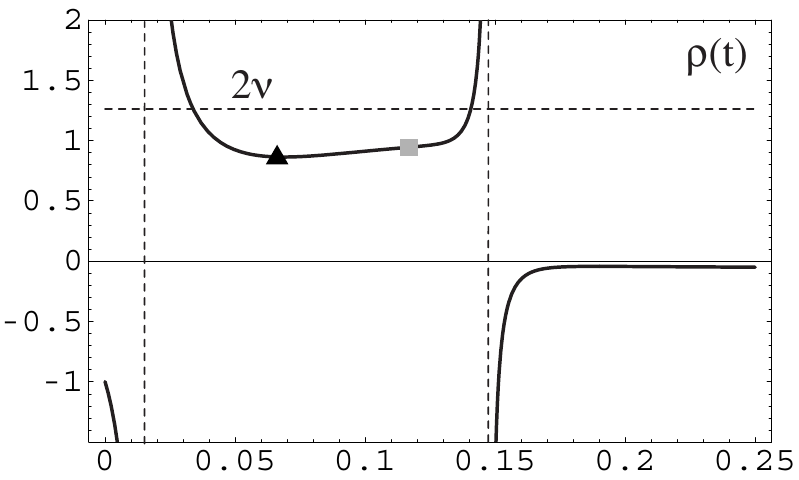}
\includegraphics[scale=0.65]{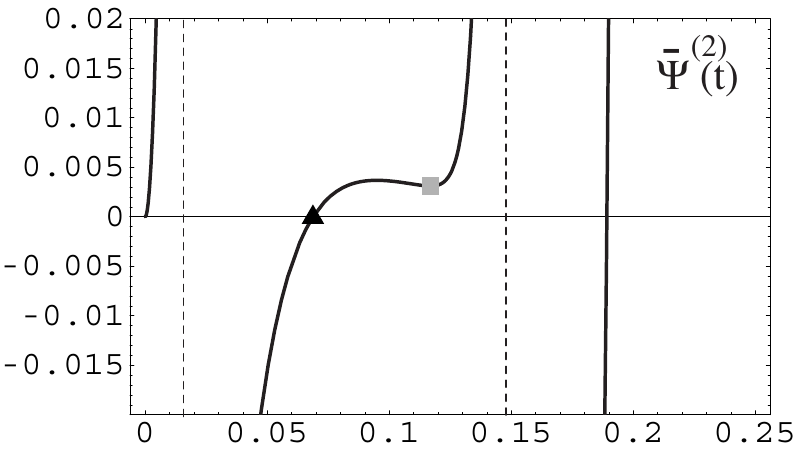}
\includegraphics[scale=0.65]{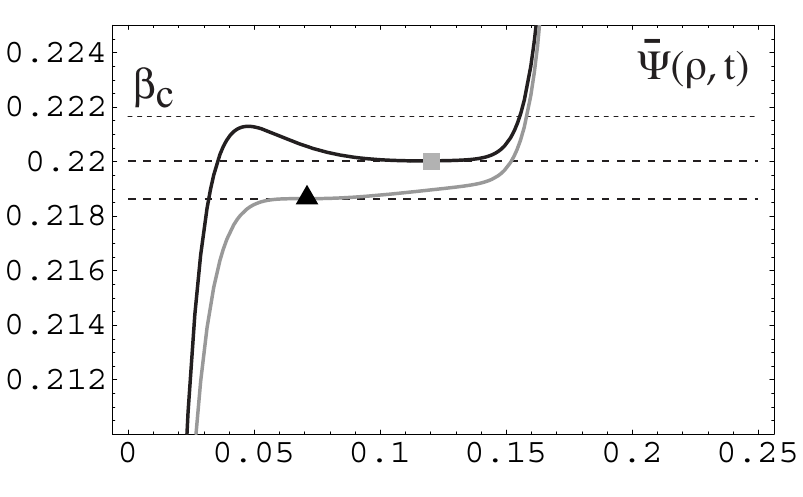}
\caption{$\rho(t)$, $\bar\psi_{N}^{(2)}(\rho(t), t)=((\bar\beta_{N}^{(2)})^2-\bar\beta_{N}^{(1)}\bar\beta_{N}^{(3)})/\bar\beta_{N}^{(2)}$ and $\bar\psi_{N}(\rho^{*}, t)$ at $N=13$.  The singular point in the upper two graphs comes as the solution of $\bar\beta^{(2)}=0$.  They are at $t=0.01526$ and $0.14768$. Note that between the two points, $\rho$ is positive definite.  The range of optimal stationary points are thus limited within $0.01526 <t^{*}< 0.14768$.  We notice that the largest solution of $\bar\psi_{13}^{(2)}=0$ exists at $t=0.18924$.  As is seen in the $\rho(t)$ plot,  this however leads to negative $\rho$ and negative $p_{1}^{-1}$.  Hence, we neglect it.  Two candidates of physical interest are shown by the points indicated by black triangle and gray box.  Among them, important one is at larger $t$ because it lies in the place where the critical region begins to appear.  The lower plot shows two $\bar\psi_{13}(\rho^{*}, t)$.  The gray one is for $\rho^{*}=0.863705$ and the black one is for $\rho^{*}=0.945108$.}
\end{figure}
The parameter $\rho$ given as a function of $t$ as ({\ref{rho3}) should behave at $t$ in the critical region as
\begin{eqnarray}
\rho&=&\frac{\sum_{k=1}^{\infty} p_{k} C(N,-p_{k})A_{k}t^{-p_{k}}}{\sum_{k=1}^{\infty}(p_{k})^2C(N,-p_{k})A_{k}t^{-p_{k}}}\nonumber\\
&=&\frac{1}{p_{1}}-\frac{p_{2}C(N, -p_{2})A_{2}}{p_{1}^3 C(N, -p_{1})A_{1}}(p_{2}-p_{1}) t^{-(p_{2}-p_{1})}+\cdots.
\end{eqnarray}
Then $\bar\psi_{N}^{(2)}$ loses $t^{-p_{1}}$ term and has to behave as 
\begin{equation}
\bar\psi_{N}^{(2)}(\rho(t),t)=\frac{p_{2}}{p_{1}}(p_{1}-p_{2})^2 C(N, -p_{2})A_{2}t^{-p_{2}}+\cdots.
\end{equation}
Note that the correction vanishes in the $N\to \infty$ limit, since $C(N, -p_{2})\to 1/\Gamma(1-p_{2}) \times N^{-p_{2}}$ as $N\to \infty$.   At finite order the correction $\sim const\times t^{-p_{2}}$ remains and, therefore, we understand that the weak peak about $t\sim 0.09$ at $N=13$ shows the transition from high temperature to scaling regions.   This is supported by the higher order plots of $\bar\psi_{N}^{(2)}$ (see the right graphs in Figure 3).  The dotted vertical lines in the left and right graphs in Figure 2 represent the value of $t$ at which $\bar\beta_{13}^{(2)}=0$ and $\rho\to \pm \infty$.  

As shown in Figure 2, at $13$th order, we obtain two candidates of the estimated $\beta_{c}$.  One comes from the zero of  $\bar\psi_{13}^{(2)}$ and the other from the local minimum of  $|\bar\psi_{13}^{(2)}|$ but $\bar\psi_{13}^{(2)}\neq 0$.  The first candidate lies in the pre-scaling region and the other in the beginning of the scaling region.  Hence, we regard the candidate at lager $t$ as important.  Actual values computed are $\beta_{c}(13)=0.218638\, ({\rm at}\,\,t^{*}=0.068789)$ and $\beta_{c}(13)=0.220024\, ({\rm at}\,\,t^{*}=0.116908)$.  As we guessed, we have thus confirmed that the best choice is the one at larger $t^{*}$.  We note that at the point where $\bar\psi_{N}^{(2)}=0$, $\rho(t)$ becomes extremal.  At general order $N$, this is found by the direct differentiation of $\rho(t)$ by $t$.
\begin{figure}[ht]
\centering
\includegraphics[scale=0.65]{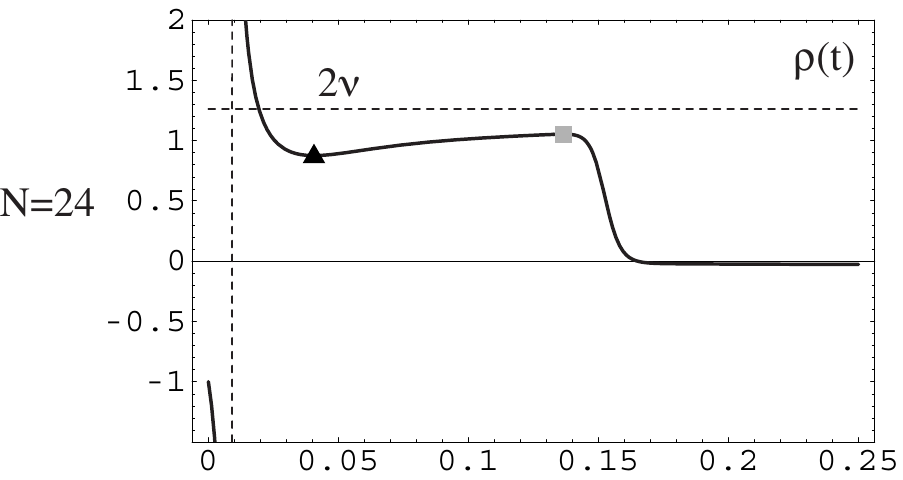}
\includegraphics[scale=0.65]{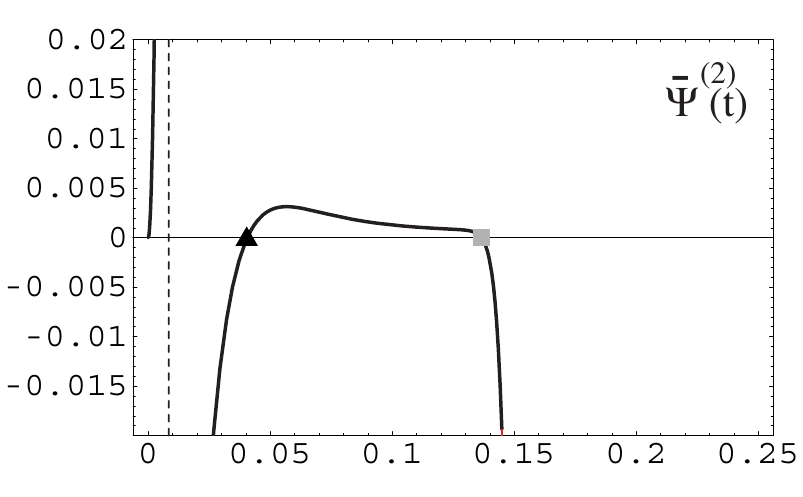}
\includegraphics[scale=0.65]{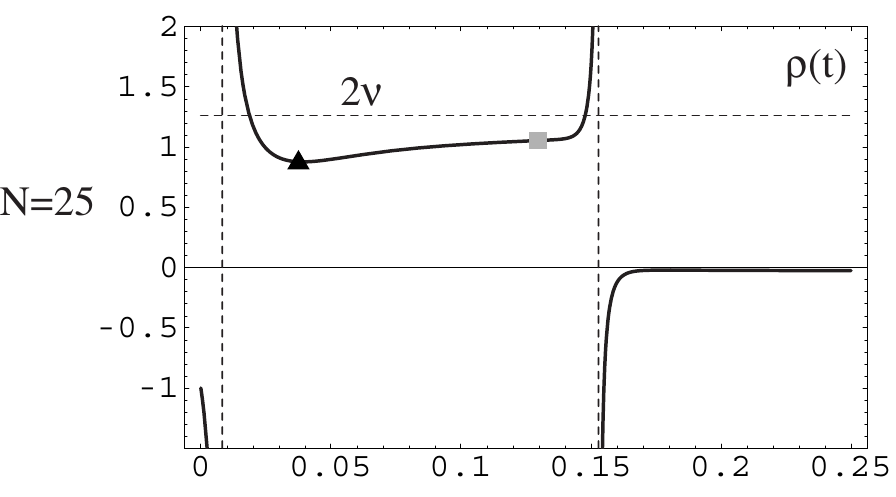}
\includegraphics[scale=0.65]{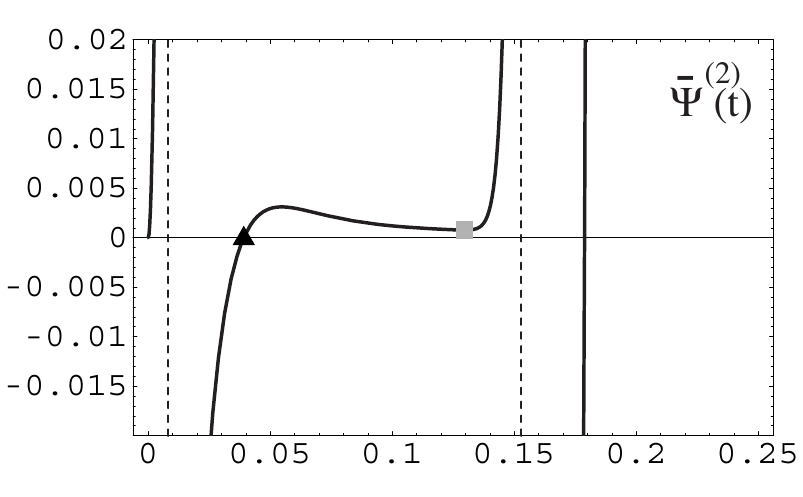}
\caption{$\rho(t)$ (left) and $\bar\psi_{N}^{(2)}(t)=((\bar\beta_{N}^{(2)})^2-\bar\beta_{N}^{(1)}\bar\beta_{N}^{(3)})/\bar\beta_{N}^{(2)}$ (right) at $N=24$ (upper) and $N=25$ (lower).  The zero point of $\bar\psi_{N}^{(2)}(t)$ at smallest $t>0$ moved toward left compared to the $13$th order.   Thus, both scaling regions have developed broader compared to the $13$th order.  This agrees with the assumption (\ref{delpsilimit}).  At $25$th order, largest solution of $\bar\psi_{25}^{(2)}=0$ exists at $t=0.178826$.  This however leads to negative $\rho$ and negative $p_{1}^{-1}$.  Hence, we neglect it.}
\end{figure}
 
Then, let us turn to the high order cases, $24$th and $25$th orders.  From the plots shown in Figure 3, we find longer and clearer scaling behavior than $13$th order.  At $24$th order, it is sufficient to confine ourselves with the region $\rho>0$.   Though there are two zeroes of $\bar\psi_{24}^{(2)}$, we may select the solution at larger $t$.   At $25$th order, the feature is similar with that at $13$th order.  The difference, however, consists in the clearness of the scaling region after the weak peak.   This is because zero of $\bar\psi_{25}^{(2)}$ has moved toward the origin and the scaling region has developed.  Due to the same reason with $13$th order case, we should rely upon the estimation at larger $t$.  In this manner, we can identify the good sequence of estimated $\beta_{c}$.  The results from $20$th to $25$th orders are shown in Table 1.  In Figure 4, the sequence of $\beta_{c}$ estimated at smaller $t$ is shown in black triangles and that at larger $t$, consisting good sequence, by gray boxes. \begin{table}
\caption{Estimation of $\beta_{c}$ and $p_{1}=1/(2\nu)=0.79365$ ($1/p_{1}\sim1.26$) with one parameter.}
\begin{center}
\begin{tabular}{ccccccc}
\hline\noalign{\smallskip}
$order$  & 20 & 21 & 22 &  23 & 24 & 25   \\
\noalign{\smallskip}\hline\noalign{\smallskip}
$\beta_{c}$   &  0.220933    & 0.220944    & 0.221035 &  0.221040 & 0.221114  & 0.221117\\
$1/p_{1}$   & 1.028607    &  1.029841    & 1.041745 & 1.042425 &  1.053098   & 1.053408\\
$ t^{*} $  &  0.133502    &  0.126393   &  0.135082  &  0.128043 &  0.136505  &  0.129586 \\
\noalign{\smallskip}\hline
\end{tabular}
\end{center}
\end{table}
\begin{figure}[ht]
\centering
\includegraphics[scale=0.65]{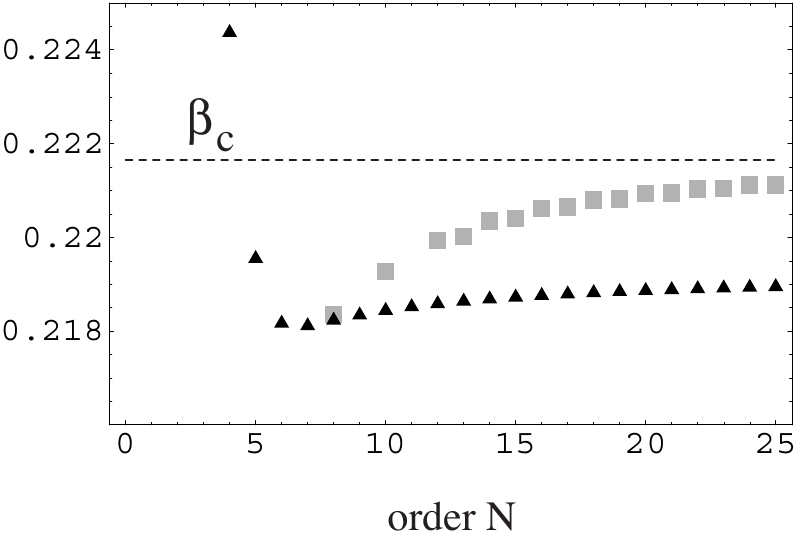}
\includegraphics[scale=0.65]{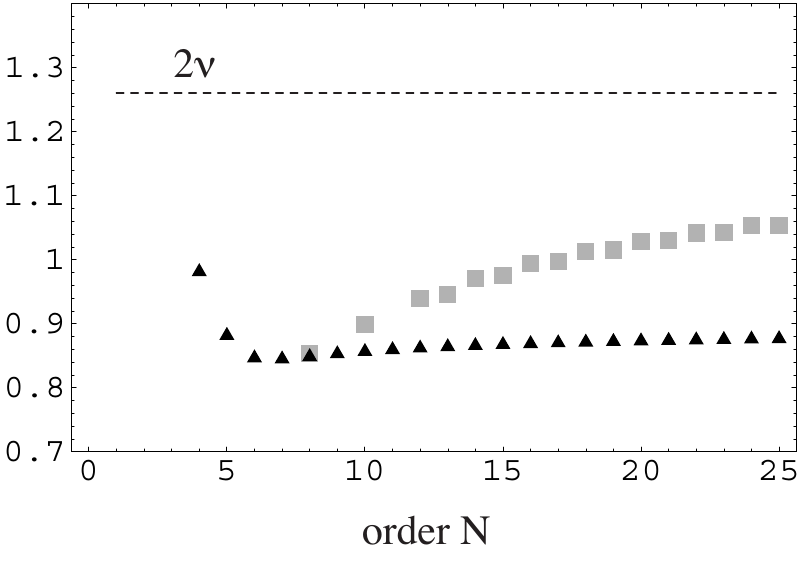}
\caption{Estimated result of $\beta_{c}$ and $p_{1}^{-1}=2\nu$ from $4$th to $25$th orders.  Black triangle indicates the result at smaller $t$ and the gray box the result at larger $t$.  The channel of good sequence opens at $8$th order for $N$ and $13$th order for odd $N$.}
\end{figure}

As the order increases, the optimized value of $\rho$ gradually increases but not to be seen to converging the established value, $1/p_{1}=2\nu=1.26$.  On the other hand, the accuracy of estimated $\beta_{c}$ is good for $N=25$ as $\beta_{c}=0.221117\cdots$ and the relative error is about $0.24$ \%.    The non-accurate results for $p_{1}$ may be explained as follows:  In our method, the parameter $\rho$ is selected for the total corrections to be suppressed.  Here notice that the dominant part of total corrections are composed by first few terms in (\ref{delbetascaling}) and not only by $C(N, -p_{1})A_{1}t^{-p_{1}}$.  Hence, even when the cancelation is successful, it does not necessarily mean that the sole information of $p_{1}$ is accurately absorbed in $\rho^{*}$, unless the order of $1/M$ expansion is extremely large.  

The sequence of good estimates starting with $8$th order for even $N$ and $13$th for odd $N$ has enough number of terms to do extrapolation to the infinite order by the fitting of obtained estimation.  Assuming the simplest  form, $\beta_{c}(N)\sim \beta_{c}-b\times N^{-u}$, let us use the last three estimations of odd orders, $21$st, $23$rd and $25$th.  Then, we obtain
\begin{equation}
\beta_{c}=0.221642,
\label{extrapolation}
\end{equation}
with $b=0.101171$ and $u=1.634371$.  If we use the $20$th, $22$nd and $24$th orders results, we obtain $\beta_{c}=0.221624$ with $b=0.10283$ and $u=1.669628$.   Both values come to close to the established value, $\beta_{c}=0.22165$  (see (\ref{betac})).  The relative error is only $0.00035\sim 0.01 $ \%.   Also for $p_{1}^{-1}$ we have done the fitting and obtained with $21$st, $23$rd and $25$th order results,
\begin{equation}
p_{1}^{-1}\approx 1.2828.
\end{equation}
And with $20$th, $22$nd and $24$th orders, we have $p_{1}^{-1}\approx 1.2640$.  
The estimation is much improved as having relative discrepancy to (\ref{nu}) about $0.3\sim 1.8$ \%.  We can say that our extrapolation provides consistent values with the present world standard.

\subsection{$\beta_{c}$ and $\nu$ in two-parameters case}
 In this subsection, we extend our method by introducing two parameters $\rho$ and $\sigma$ associated with $\beta^{(1)}$ and $\beta^{(2)}$, respectively.   We expect that the problem of the inaccuracy of estimated $p_{1}$ may be partially resolved by introducing multi-parameters.
 
Consider the function $\psi(\rho, \sigma, x)$ which has the same limit as $x\to \infty$ with $\beta$,
\begin{equation}
\psi(\rho, \sigma, x)=\Big[1+\rho x\frac{d}{dx}+\sigma \Big(x\frac{d }{dx}\Big)^2\Big]\beta.
\end{equation}
The delta expansion of $\psi_{N}$ reads
\begin{equation}
\bar\psi_{N}(\rho, \sigma, t)=\Big[1+\rho t\frac{d}{dt}+\sigma \Big(t\frac{d }{dt}\Big)^2\Big]\bar\beta_{N}(t)=\bar\beta_{N}+\rho\bar\beta_{N}^{(1)}+\sigma\bar\beta_{N}^{(2)}.
\end{equation}
Under some domain of $(\rho, \sigma)$ and some region of $t$, we postulate
\begin{equation}
\lim_{N\to \infty}\bar\psi_{N}(\rho,\sigma,t)=\beta_{c}.
\end{equation}
Then, since $\bar\psi_{N}$ behaves at large $t$ as,
\begin{equation}
\bar\psi_{N}(\rho, \sigma, t)=\beta_{c}-\sum_{k=1}^{\infty}(1-\rho p_{k}+\sigma p_{k}^2)C(N,-p_{k})A_{k}t^{-p_{k}},
\label{dif2}
\end{equation}
we infer that, at least for large enough $N$, optimal $\rho$ and $\sigma$ which realize the maximal stationarity may lead that the first two corrections in (\ref{dif2}) effectively disappears.  It then follows that
\begin{equation}
1-\rho p_{1}+\sigma p_{1}^2\sim 0,\quad 1-\rho   p_{2}+\sigma p_{2}^2\sim 0.
\label{p1p2}
\end{equation} 
We use these upon the estimation of $p_{i}$ from optimal $\rho$ and $\sigma$.

Since we have two free parameters, we impose stationarity criteria up to the second derivative,
\begin{equation}
\bar\psi_{N}^{(1)}(\rho, \sigma, t)=0,\quad \bar\psi_{N}^{(2)}(\rho, \sigma, t)=0.
\end{equation}
In terms of $\beta$, the above condition reads as
\begin{eqnarray}
\bar\beta_{N}^{(1)}+\rho \bar\beta_{N}^{(2)}+\sigma \bar\beta_{N}^{(3)}&=&0,\nonumber\\
\bar\beta_{N}^{(2)}+\rho \bar\beta_{N}^{(3)}+\sigma \bar\beta_{N}^{(4)}&=&0.
\label{2para}
\end{eqnarray}
For a given set of $(\rho, \sigma)$, we search for the stationary points at which (\ref{2para}) is satisfied.    As in the one-parameter case, estimated $\beta_{c}$ varies with values of $(\rho, \sigma)$ and we have to extract the optimal set among them.  The condition is that, just at $t$ which is the solution of (\ref{2para}), $|\bar\psi_{N}^{(3)}|$ becomes local minimum.   To proceed further, it is convenient to express $\rho$ and $\sigma$ in terms of the solution $t$ for (\ref{2para}).  It is then readily obtained that
\begin{eqnarray}
\rho&=&\frac{\bar\beta_{N}^{(2)}\bar\beta_{N}^{(3)}-\bar\beta_{N}^{(1)}\bar\beta_{N}^{(4)}}{\Delta},\\
\sigma&=&\frac{\bar\beta_{N}^{(1)}\bar\beta_{N}^{(3)}-(\bar\beta_{N}^{(2)})^2}{\Delta},
\label{rhosig}
\end{eqnarray}
where
\begin{equation}
\Delta=\bar\beta_{N}^{(2)}\bar\beta_{N}^{(4)}-(\bar\beta_{N}^{(3)})^2.
\label{Delta}
\end{equation}
The two parameters behave at large $t$ as
\begin{eqnarray}
\rho&\sim& \frac{p_{1}+p_{2}}{p_{1}p_{2}}+const\times t^{-p_{3}+p_{2}}+\cdots,\nonumber\\
\sigma&\sim&\frac{1}{p_{1}p_{2}}+const\times t^{-p_{3}+p_{2}}+\cdots.
\end{eqnarray}
This shows the critical behaviors of two parameters.   The critical behavior of $\bar\psi^{(3)}(t):=\bar\psi^{(3)}(\rho(t), \sigma(t), t)$ is then given by $\bar\psi^{(3)}(t)\sim const\times t^{-p_{3}}$.   The coefficient of $t^{-p_{3}}$ denoted by "const" depends on the order $N$ and tends to zero in the $N\to \infty$ limit.   Therefore, 
as in the one-parameter case, we consider the minimization of $|\bar\psi^{(3)}(t)|$ including the case,
\begin{equation}
\bar\psi_{N}^{(3)}(t)=\bar\beta_{N}^{(3)}+\rho(t) \bar\beta_{N}^{(4)}+\sigma(t) \bar\beta_{N}^{(5)}=0.
\end{equation}
Then, with obtained $t^{*}$ and $\rho^{*}=\rho(t^{*})$ and $\sigma^{*}=\sigma(t^{*})$, 
the estimation of $\beta_{c}$ is directly given by
\begin{equation}
\beta_{c}(N)= \bar\psi_{N}(\rho^{*}, \sigma^{*}, t^{*}).
\end{equation}

Figure 5 shows the plot of $\bar\psi_{25}^{(3)}$ as function of $t$ in different scales.  The vertical dotted lines indicate the singularities of $\bar\psi_{25}^{(3)}$.  All of them come from zero points of $\Delta$, the solutions of $\Delta=0$.   The stationary point with $|\bar\psi_{25}^{(3)}|$ locally minimum but non-zero lives in the left region separated by one of the lines.  In fact, the corresponding value of $\sigma$ vanishes and the solution corresponds to the one obtained in one-parameter case with {\it smaller $t^{*}$}.  Thus, it still lies in the transit region from high temperature to scaling regions.  The right region separated by the right vertical dotted line is also not physically interesting, because in the region, $\sigma$ becomes negative.  
\begin{figure}[ht]
\centering
\includegraphics[scale=0.7]{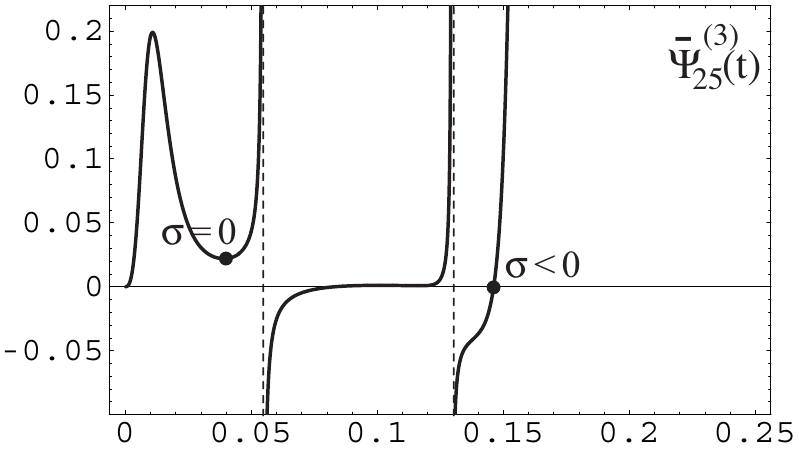}
\includegraphics[scale=0.75]{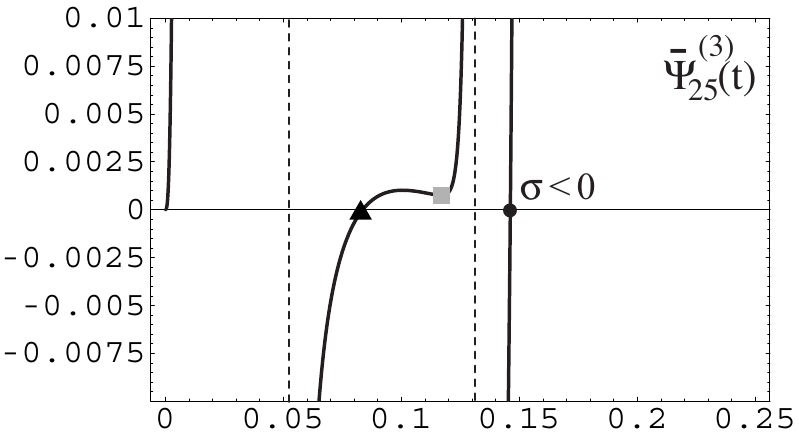}
\caption{$\bar\psi_{25}^{(3)}(t)$ with two scales.  In the left plot, the point shown in the filled circle at $t=0.039$ represents the local minimum of $\bar\psi_{25}^{(3)}$ and leads $(\rho,\sigma)=(0.87626, 0)$.  This reduces to the one-parameter case.  The filled circle at $t=0.146$ gives negative $\sigma$ and we neglect the solution.   In the right plot, we emphasized two candidates both of which provide positive two parameters.  The gray box indicates the element of good sequence of $\beta_{c}$ estimation.}
\end{figure}
Even if we would admit small negative $\sigma$ for the compensation of a bit too large $\rho$-value, 
 (\ref{p1p2}) gives
\begin{eqnarray}
\rho&=&\frac{1}{p_{1}}+\frac{1}{p_{2}},\\
\sigma&=&\frac{1}{p_{1}p_{2}},
\end{eqnarray}
and negative $\sigma^{*}$ leads to negative $p_{2}$ for $p_{1}>0$, which contradicts to $0<p_{1}<p_{2}<\cdots$.   Hence, the case of negative $\sigma^{*}$ does not provide us truly reliable estimation of $p_{1}$ and $p_{2}$.   From these consideration, we conclude that the scaling region is roughly implied by the two dotted vertical lines and only the cases $\sigma^{*}>0$ are worth of serious consideration.  In the right plot in Figure 5, two such solutions labelled by black triangle and gray box are shown.

\begin{table}
\caption{Estimation of $\beta_{c}$, $p_{1}$ and $p_{2}$ with two parameters.}
\begin{center}
\begin{tabular}{ccccccc}
\hline\noalign{\smallskip}
$order$ & 20 & 21 & 22 & 23 & 24 & 25   \\
\noalign{\smallskip}\hline\noalign{\smallskip}
$\beta_{c}$ &  0.221442  & 0.22140 & 0.221505 &   0.221513  &  0.221553  & 0.221562 \\
$1/p_{1}$ & 1.138882 & 1.127291 & 1.160289  & 1.163069  &  1.179823  & 1.183677   \\
$1/p_{2}$ & 0.433345 & 0.413636  & 0.468705  & 0.472926  & 0.500057 &  0.505965\\
$ t^{*} $  & 0.116906 &  0.096589  & 0.119071 &  0.114190  &  0.120643  &  0.116991 \\
\noalign{\smallskip}\hline
\end{tabular}
\end{center}
\end{table}

\begin{figure}[h]
\centering
\includegraphics[scale=0.7]{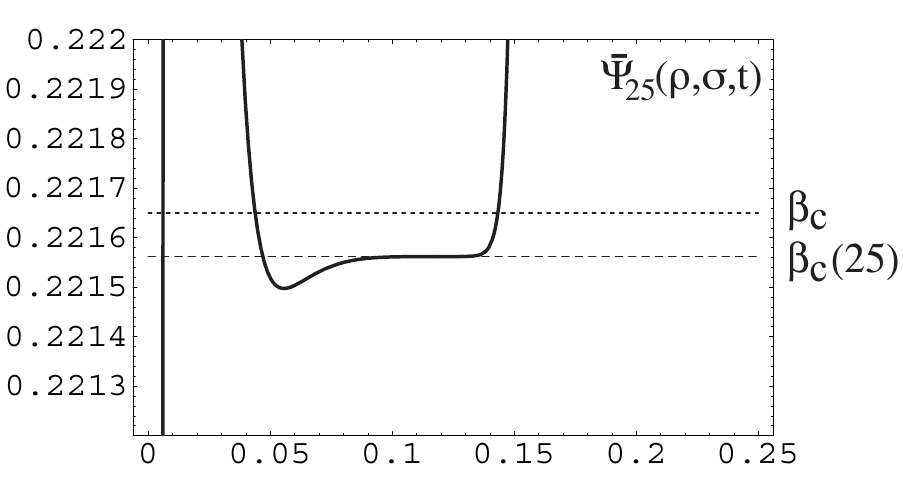}
\caption{$\bar\psi_{25}(\rho, \sigma, t)$ for $\rho=1.6896423$, $\sigma=0.5988992$.  The two dotted lines indicate $\beta_{c}=0.22165$ and its estimation, $\beta_{c}(25)=0.221562$.}
\end{figure}

\begin{figure}[h]
\centering
\includegraphics[scale=0.75]{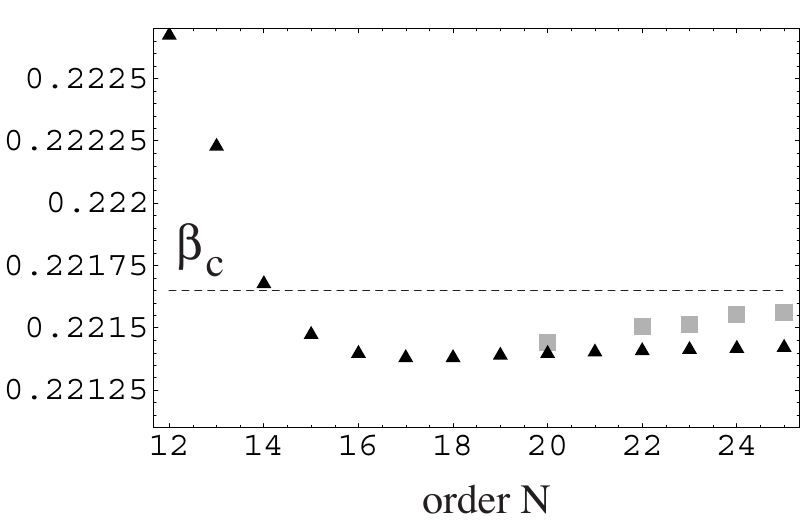}
\includegraphics[scale=0.7]{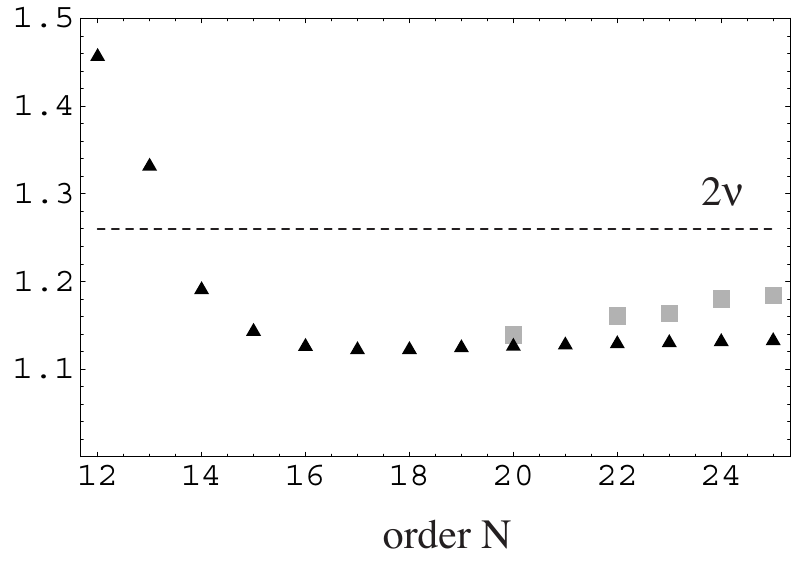}
\caption{Plots of estimated $\beta_{c}$ and $p_{1}^{-1}=2\nu$ with two-parameters.  The dotted lines indicate $\beta_{c}=0.22165$ and $2\nu=1.26$.  The gray boxes indicate accurate sequence of which elements appears from $20$th order.}
\end{figure}
The result of estimation is shown in Table 2, Figure 6 and Figure 7.   The behavior of $\bar\psi_{N}^{(3)}$ becomes steady at $12$th order and single solution is obtained up to $19$th order.  The new channel to the most accurate sequence opens from $20$th for even orders and from $23$rd for odd orders.  This sequence includes estimation lying in the inside of scaling region.  The level of the scaling behavior observed in $\bar\psi_{25}^{(3)}$ is, however weak compared to the one-parameter case.  The scaling level of $25$th order in two-parameter case is, to the eye, the same level with the $13$th order in one-parameter case.   However the accuracy is improved.  The relative error of $25$th order $\beta_{c}$ is about $0.004$ \%.  On the other hand, the estimation of $p_{1}^{-1}=2\nu$ is not so good, though two-parameter estimation is improved compared to one-parameter case.  At $25$th order, the relative error is about $6$ \%.

The extrapolation of the good sequence to the $N\to \infty$ limit is not adequate in two-parameter case.  This is because the number of elements in the sequence is not enough. 

\subsection{$\beta_{c}$ and $\nu$ in three-parameters case and more}
With three parameters $\rho$, $\sigma$ and $\tau$, we deal with
\begin{equation}
\psi(\rho,\sigma,\tau;x)=\Big[1+\rho x\frac{d}{dx}+\sigma^2(x\frac{d}{dx})^2+\tau(x\frac{d}{dx})^3\Big]\beta.
\end{equation}
The estimation procedure follows those of one- and two-parameter cases and we omit the details and present just the outline.  The stationarity condition reads $\bar\psi_{N}^{(i)}=0$, $(i=1,2,3)$ and then $\rho$, $\sigma$ and $\tau$ are given by the following equations as function of $t$,
\begin{equation}
\bar\beta_{N}^{i}+\rho\bar\beta_{N}^{(i+1)}+\sigma\bar\beta_{N}^{(i+2)}+\tau\bar\beta_{N}^{(i+3)}=0,\quad (i=1,2,3).
\end{equation}
Then, consider the minimization of $|\bar\psi_{N}^{(4)}|$ and the determination of $t^{*}$ gives optimal set $(\rho^{*}, \sigma^{*}, \tau^{*})$ and
\begin{equation}
\beta_{c}(N)=\bar\psi_{N}(\rho^{*},\sigma^{*},\tau^{*},t^{*}).
\end{equation}
Estimation of $p_{i}\,(i=1,2,3)$ can be given by $(\rho^{*},\sigma^{*},\tau^{*})$ via
\begin{eqnarray}
\rho&=&\frac{1}{p_{1}}+\frac{1}{p_{2}}+\frac{1}{p_{3}},\nonumber\\
\sigma&=&\frac{1}{p_{1}p_{2}}+\frac{1}{p_{2}p_{3}}+\frac{1}{p_{3}p_{1}},\nonumber\\
\tau&=&\frac{1}{p_{1}p_{2}p_{3}}.
\end{eqnarray}

With three parameters, $\bar\psi_{N}^{(4)}$ shows complicated behavior to $22$th order.   Only from $23$th order, our method begins to provide estimations characteristic to the three-parameter case.  Even then, the behavior of $\bar\psi_{N}^{(4)}$ is not matured yet compared to the higher orders in the one- and two-parameter cases.  It is typically reflected to the narrowness of the scaling region where all of $\rho$, $\sigma$ and $\tau$ are positive.  In addition, at the last even order $N=24$ in three-parameter case, the behavior of $\psi_{24}^{(4)}$ resembles to the $12$th or $14$th orders in the two-parameter case.  See the left plot in Figure 8.   Table 3 shows the results at $23$rd, $24$th and $25$th orders.   As results in the first several orders exceeded of the correct value of $\beta_{c}$ in one- and two-parameter cases, these three estimations exceed $\beta_{c}=0.22165$, though the last order estimation is most close to the established value.  We notice that at $25$th order, the estimation of $p_{1}^{-1}$ is also most accurate.  This implies that when $\beta_{c}$ is precisely obtained, estimated $\nu$ is also accurate.   The opening of the accurate sequence for the three-parameter case demands further computation of high temperature expansion, maybe up to $30$th order or more.   

\begin{table}
\caption{Estimation of $\beta_{c}$ and $p_{1}$ and $p_{2}$ with three parameters.}
\begin{center}
\begin{tabular}{cccc}
\hline\noalign{\smallskip}
$order$ &  23 &  24 & 25    \\
\noalign{\smallskip}\hline\noalign{\smallskip}
$\beta_{c}$  & 0.222079  & 0.221741 & 0.221687   \\
$1/p_{1}$ & 2.010008 & 1.336976 & 1.274452   \\
$1/p_{2}$ &0.920863 & 0.715954  & 0.648486   \\
$1/p_{3}$ &0.299087 & 0.242590  & 0.210993  \\
$ t^{*} $  & 0.111805 & 0.111116  & 0.112973  \\
\noalign{\smallskip}\hline
\end{tabular}
\end{center}
\end{table}

\begin{figure}[h]
\centering
\includegraphics[scale=0.75]{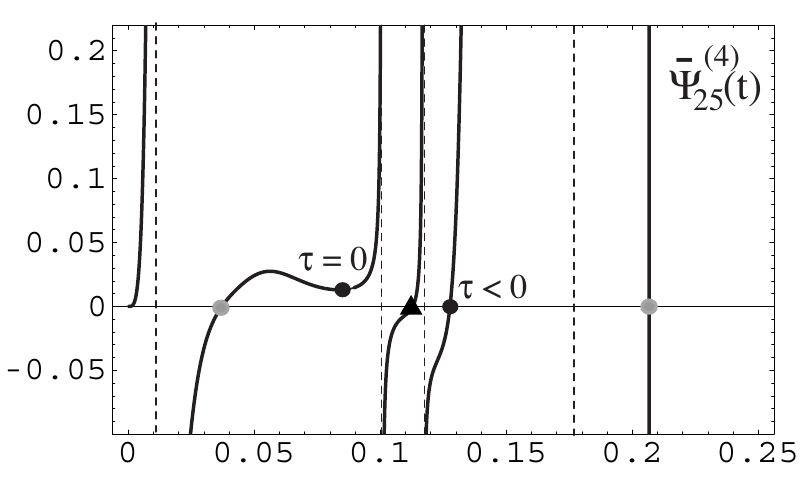}
\includegraphics[scale=0.75]{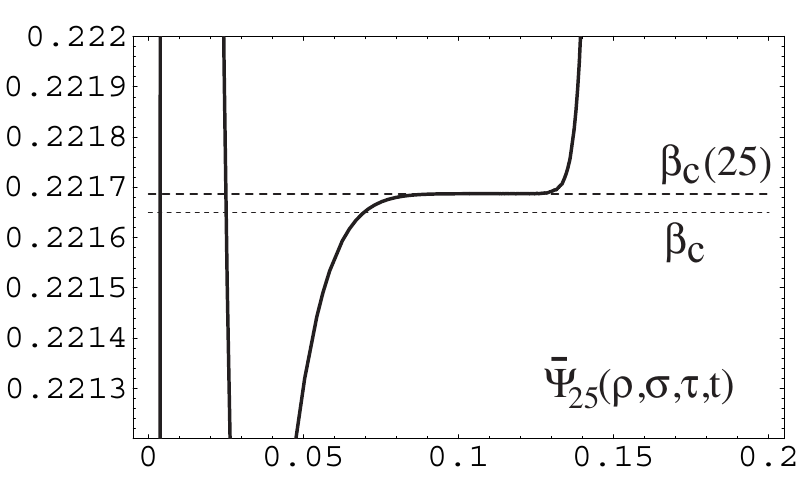}
\caption{ Left plot: $\bar\psi_{25}^{(4)}(t)$.  There are five points at which $|\bar\psi_{25}^{(4)}(t)|$ becomes locally minimum.  The left gray circle gives positive $\rho$, $\sigma$ and $\tau$.  However, the two of $p_{i}$ $(i=1,2,3)$ becomes complex conjugates.  The two black circles give $\tau=0$ and $\tau<0$ solution, respectively from the left to the right.  The solution $\tau=0$ corresponds to the solution in the $2$-parameter case with {\it smaller $t=0.0837$} (Represented by a black triangle in the right plot in Figure 5.  This solution lies outside of scaling region and not included in the good sequence.)  The rightest candidate at $t=0.2067$ gives two of $p_{i}$ $(i=1,2,3)$ negative and we neglect it.  Thus, the possible choice is only the black triangle for which $\beta_{c}$ and $p_{i}$ give values shown in Table 3.   The region where the black triangle included is too narrow and this implies that the order of expansion is still short for clear demonstration of the scaling for $3$-parameter $\bar\psi_{N}$.   Right plot:  $\bar\psi_{25}(\rho,\sigma,\tau, t)$ for $\rho=2.1339308$, $\sigma=1.2321906$ and $\tau=0.1743783$.  The dotted line indicates $\beta_{c}(25)$ to the established digits, $\beta_{c}=0.22165$.}
\end{figure}

Under the simple estimation without using extrapolation to the infinite order, increasing the number of free parameters improves the estimation of critical quantities so far.  However, the reliable estimation with confidence of the scaling behavior of relevant functions sets in larger orders when the number of parameters are increased.  This stems from the fact that the differentiation on $\bar\beta_{N}(t)$ creates oscillation and delays the appearance of scaling behavior.  To say in the detailes the rationale is as follows:  The small $t$ behavior of $\bar\beta_{N}^{(k)}(t)$ reads
\begin{equation}
\bar\beta_{N}^{(k)}(t)\sim -\sum_{n=1}^{\infty}A_{n}C(N, -p_{n})(-p_{n})^k t^{-p_{n}}.
\end{equation}
From $0<p_{1}<p_{2}<p_{3}<\cdots$ and the result in Table 3, we find $p_{2}>1$.  Then for $n\ge 2$,  $p_{n}^k$ $(n\ge 2)$ grows with $k$.   This means that the differentiation enhances higher order corrections and the critical behavior of $\bar\beta_{N}^{(k)}$ is obscured.   While at small $t$, the differentiation on $\bar\beta$ creates $n$ to the coefficient of $t^{n}$.  Hence, also in small-$t$ expansion, the upper limit of effective region of $\bar\beta_{N}^{(k)}$ tends to shrink.  We have actually found that, up to $25$th order, $m$-parameter extension when $m\ge 4$ does not work.    
In our study up to $25$th order, the three-parameter case is at the limit of multi-parameter extension of $\beta$ being effective.

\section{Discussion and Conclusion}
It would be better to mention on the $p_{2}$ estimation.  As would be understood from Table 2 and Table 3, the estimation of $p_{2}$ is not successful in our approach.  If one uses standard values for $\nu=0.63$ and $\theta=0.5$, one has $p_{2}^{-1}\sim 0.84$.  Our estimated results in two- and three- parameters are still far from the value.   Perhaps, $p_{2}$ and also $p_{1}$, may be obtained with more accuracy if we can invent the method where we directly address to $p_{i}$ estimation.  In our approach, parameters $\rho$, $\sigma$ and $\tau$ are just optimized in order to estimate $\beta_{c}$. 

Now, let us give another comment.  The behaviors of the sequences of $\beta_{c}$ and $\nu$ share among themselves a  similar pattern according to the expansion order $N$.   The pattern is such that the estimated result decreases first and, after the bottom out, turn to increase and then the good sequence appears (see Figure 4 and Figure 7).   Also when the number of parameters was increased from single to double, same pattern repeated though the onset of the pattern delayed for several orders.     The three parameter case would also follow this course.  On the sequence of $\beta_{c}(N)$ first appearing, we note that any element stays within pre-scaling region.  It might shoot wrong value in the $N\to \infty$ limit.   In fact, the extrapolation in the simple ansatz used in section \ref{sec:one-para} yields value plagued with non-negligible discrepancy with the correct value.   Presumably the three parameter case is not exceptional, too.

With focus on the scaling behavior of $\psi\sim \sum_{k=0}\rho_{k}\beta^{(k)}$, we have attempted the estimation of $\beta_{c}$ and $\nu$.  In the one- and two-parameter cases, good sequence emerged in the timing that  the scaling regions of all relevant functions including $\psi$, $\psi^{(1)}$, $\cdots$ begin to appear.  That sequence is most important since the element lies in the deepest place of the observed scaling region.  The accuracy of the estimated results of our approach is not so superior compared with other approaches explained in~\cite{peli}.   Especially, the estimation of $\nu$ needs more improvement for the higher accuracy.  This remains as a problem of our approach.  However, in our approach, it is transparent that which one is important when two or more candidates are present at an order.  It becomes possible because the critical behaviors in relevant functions can be directly visible.  At a fixed order, we can thus find single best estimation systematically, which is desirable for theoretical approach.


\end{document}